\documentclass[12pt,a4paper]{article}
\usepackage[utf8]{inputenc}
\usepackage{amsmath}
\usepackage{amsfonts}
\usepackage{amssymb}
\usepackage{verbatim}
\usepackage{mathrsfs}
\usepackage{amsthm}

\usepackage[dvipsnames,svgnames,x11names,hyperref]{xcolor}

\usepackage{enumitem}
\setlist[enumerate]{nosep}
\setlist[enumerate]{topsep=0pt}

\theoremstyle{definition} 
\newtheorem{theorem}{Theorem}
\newtheorem{theorema}{Theorem}
\newtheorem{lemma}{Lemma}
\newtheorem{definition}{Definition}

\usepackage{setspace}
\DeclareMathOperator*{\argmax}{arg\,max}

\usepackage{tikz}
\usetikzlibrary{trees}

\tikzstyle{level 1}=[level distance=2.5cm, sibling distance=3.5cm]
\tikzstyle{level 2}=[level distance=2.5cm, sibling distance=3cm]
\tikzstyle{level 3}=[level distance=3cm, sibling distance=2.5cm]

\tikzstyle{decision} = [rectangle, minimum height=12pt, minimum width=12pt, draw=black, fill=none]
\tikzstyle{chance} = [circle, minimum width=12pt, draw=black, fill=none]
\tikzstyle{outcome} = [circle, minimum width=2pt,fill, inner sep=0pt]

\usepackage[backend=bibtex, style=authoryear-comp, sorting=nyt, sortcites=false, maxbibnames=10]{biblatex}
\usepackage{hyperref}

 \hypersetup{
 	pdfauthor={Sven Neth},
	pdftitle={Rational Aversion to Information},
	pdfkeywords={Value of Information, Decision Theory, Bayesian Epistemology. Probability},
	pdflang={en},
	hidelinks
}

\author{}
\title{Rational Aversion to Information}
\author{Sven Neth \\ Forthcoming in \emph{The British Journal for the Philosophy of Science}\thanks{See \url{https://doi.org/10.1086/727772} for final version.}}
\date{September 2023}
\begin{document}

\maketitle

\begin{abstract}
Is more information always better? Or are there some situations in which more information can make us worse off? \textcite{Good1967} argues that expected utility maximizers should always accept more information if the information is cost-free and relevant. But Good's argument presupposes that you are certain you will update by conditionalization. If we relax this assumption and allow agents to be uncertain about updating, these agents can be rationally required to reject free and relevant information. Since there are good reasons to be uncertain about updating, rationality can require you to prefer ignorance.
\end{abstract}

\section{Introduction}

We care about learning the truth for its own sake, but we also care about learning because it can lead us to make better decisions. That is, besides the epistemic benefits of finding out the truth, learning often comes with instrumental benefits as well. 

Is more information always instrumentally better? Or are there situations in which more information can make us foreseeably worse off? It is clear that information can make us worse off if we consider the cost of processing and storing the information or the opportunity cost of thinking for too long before acting. Nobody thinks that you have to read all the reviews before buying a new vacuum cleaner or that you should think long before hitting the brakes when a red light comes up. It is also clear that information can make us worse off if it is false, so let me be clear that when I talk about information, I always mean true information.

What if the information is cost-free? For rational agents, is it always instrumentally valuable to accept free information? \textcite{Good1967} argues that the answer is `yes' if we accept the principle of maximizing expected utility. However, Good presupposes that you are certain you will update by conditionalization, which means you are certain your new credences after learning are equal to your old conditional credences given the learned event. There are good reasons to assign positive probability to failures of conditionalization, even for rational agents. I show that if you assign a positive probability to failures of conditionalization, the principle of maximizing expected utility can require you to reject free information. Sometimes, even expected utility maximizers are better off knowing less. Moreover, this offers a vindicating explanation of why people sometimes reject information in real-life examples, such as medical testing.

To be clear, this paper is not about situations in which you actually fail to conditionalize. In all my examples below, we can assume that the agent conditionalizes in the actual world. Rather, this paper is about situations in which you fail to be certain that you will conditionalize. You can fail to be certain that you will conditionalize even if you always conditionalize.

Here is the plan. First, I explain Good's argument. Then, I explain why Good's argument presupposes that you are certain you will update by conditionalization and give reasons to reject this assumption. I show how assigning a positive probability to failures of conditionalization can make it rational to reject free information for expected utility maximizers and sketch how this can explain information aversion in the real world. I finish by explaining how we can generalize the value of information to agents who are uncertain about how they will update.

\section{Good's Argument}

I start by introducing some terminology and explain Good's argument.

\subsection{Terminology}\label{terminology}

I use the framework of \textcite{Savage1972} to model decision making under uncertainty. We have a set $\Omega$ of \text{states}, which contains all epistemically possible worlds from the point of view of the agent we are modeling. Events are subsets of $\Omega$ and we model the credences of our agent by a probability function.\footnote{I assume $\Omega$ is finite and model credences as finitely additive probability function $p :\mathcal{P}(\Omega) \to [0,1]$.} We also have a set $\mathscr{O}$ of \text{outcomes}, where outcomes contain everything our agent cares about. We model our agent's preferences over outcomes by a utility function $u$ which maps outcomes to their utilities. \text{Actions} are functions from states to outcomes. I assume actions are causally and probabilistically independent of states.\footnote{\textcite{Adams1980} and \textcite{Maher1990} discuss how Good's argument fails if this assumption is relaxed, in both evidential and causal decision theory.}

Given a probability function $p$ and utility function $u$, the \text{expected utility} of action $f$ is:
\begin{equation*}
\mathbb{E}_{p}(f) = \sum_{\omega \in \Omega} p(\omega)u(f(\omega)).\footnote{$p(\omega)$ is shorthand for $p(\{\omega\})$.}
\end{equation*}
I assume your utility function remains fixed through learning.\footnote{I set aside cases in which learning leads you to change your utility function, perhaps in a `transformative experience' \parencite{Paul2014, Pettigrew2019}.} However, your credences change in response to evidence, so I relativize expected utility to a probability function.

A \text{choice set} is a set of actions among which our agent makes a decision. I assume that all choice sets are finite. Our agent maximizes expected utility if for every available choice set $\mathcal{S} = \{f_1, ..., f_n\}$, she picks an action $f_i \in \mathcal{S}$ which maximizes expected utility relative to her probability function $p$ and utility function $u$.

I model learning by an \text{evidence partition} $\mathcal{E}$ of $\Omega$. This partition contains the events our agent might learn, which are mutually exclusive and collectively exhaustive. We can think of the evidence partition as a question, for example the question whether it is sunny or rainy outside. In this case, the evidence partition contains two cells: the worlds where it is sunny outside and the worlds where it is rainy outside. Since the events in the evidence partition are live possibilities for what our agent might learn, they all have non-zero probability, so $p(E) > 0$ for all $E \in \mathcal{E}$. 

When learning event $E$ in the evidence partition, our agent updates her credences to $\mathcal{P}_{E}$. I allow our agent to be uncertain about how she will update. This means that $\mathcal{P}_{E}$ is not a particular probability function, but rather a \text{random variable} whose values can be different probability functions. (Hence the fancy typeface.) The only constraint I impose is that after learning an event, our agent is certain of that event.\footnote{Let $\Delta(\Omega)$ be the set of all probability functions $p : \mathcal{P}(\Omega) \to [0,1]$. Formally, $\mathcal{P}_{E}$ is a function from $E$ to $\Delta(\Omega)$ such that for each $\omega \in E$, $\mathcal{P}_{E}(\omega)(E) = 1$. For each $\omega \in E$, $\mathcal{P}_{E}(\omega)$ is a particular probability function.} I write $p(\cdot \mid E)$ for the credences our agent adopts after learning event $E \in \mathcal{E}$ and updating by conditionalization.\footnote{I use the standard ratio definition: $p(A \mid E) = \frac{P(A \cap E)}{p(E)}$ assuming $p(E) > 0$.}

Here is an example of our framework in action. You are at the horse track thinking about which horse to bet on. The states specify which horse will win the race (and other relevant facts) and the actions are different bets you might place. The probability function $p$ encodes your credences about different horses winning and the utility function $u$ models how much you value the outcomes of these bets, for example different amounts of money.

Imagine a charming stranger comes up to you and offers you their opinion on which horse is likely to win. Are you willing to listen? The evidence partition $\mathcal{E}$ contains different opinions the stranger might voice and $\mathcal{P}_{E}$ models how you expect to update your credences after listening. To be clear, the information you learn is not which horse is likely to win but only what the stranger is saying. The stranger might be lying or clueless. 

You need to decide: Do you want to find out what the stranger has to say or would you rather place your bet now? It is not obvious how to answer this question. On the one hand, you might listen to the stranger and ignore what they say if you do not find it helpful, so how could listening harm you? On the other hand, perhaps the stranger is trying to mislead you. In this case, do you trust yourself to listen before placing your bet?

\subsection{The Argument}

\textcite{Good1967} thinks you should listen to the stranger before placing your bet. More generally, Good argues that if you are rational, then given any choice set and evidence partition, you are never worse off by first learning the true event in the evidence partition and making your choice afterwards rather than making your choice now.\footnote{\textcite{Skyrms1990} provides a helpful overview and points out that \textcite{Ramsey1990, Savage1972} give similar arguments. \textcite{Good1967} notes that his argument is partly anticipated by \textcite[90]{Raiffa1961} and \textcite[66]{Lindley1965}. \textcite{Hosiasson1931} discusses similar ideas and cites an unpublished paper by Ramsey as inspiration. Interestingly, the argument does not work when you decide whether \text{someone else} should learn more information before making a decision \parencite{Good1974}.}  Good does not mean that more information always leads to better decisions. You might get unlucky and learn something misleading. Rather, Good argues that learning cannot make you \text{foreseeably} worse off.

\begin{figure}[h]
\begin{tikzpicture}[grow=right, sloped]
\node[decision] {}
    child {
        node[chance] {}        
            child {
                node[outcome, label=right: {Bet on horse $B$. Expected payoff: $\frac{1}{4}$}] {}
                edge from parent
                node[above] {Shine}
                node[below]  {$\frac{1}{2}$}
            }
            child {
                node[outcome, label=right: {Bet on horse $A$. Expected payoff: $\frac{1}{4}$}] {}
                edge from parent
                node[above] {Rain}
                node[below]  {$\frac{1}{2}$}
            }
            edge from parent 
            node[above] {Learn}
    }
    child {
        node[outcome, label=right: {Play it safe. Expected payoff: 0}] {}    
        edge from parent         
            node[above] {Don't learn}
    };
\end{tikzpicture}
\caption{If you care about winning, you should listen to the weather report.}
\label{fig:simple}
\end{figure}
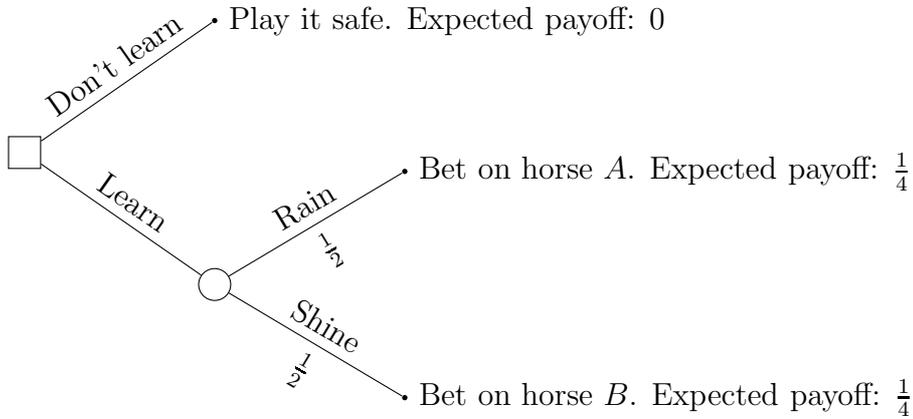

The idea behind Good's argument can be illustrated by an example. Suppose there is a race between horse $A$ and horse $B$ tomorrow. You have to decide whether (i) to bet on horse $A$, which means you win \$1 if $A$ wins and lose \$2 otherwise, (ii) to bet on horse $B$, which means you win \$1 if $B$ wins and lose \$2 otherwise, or (iii) to play it safe, which means you won't win or lose anything. You think $A$ and $B$ are equally likely to win, so your best option right now is to play it safe. But you can listen to the (accurate) weather report for tomorrow. You think that $A$ is $\frac{3}{4}$ likely to win if it rains and $B$ is $\frac{3}{4}$ likely to win if the sun shines. We can illustrate your decision problem as shown in figure \ref{fig:simple}, where rectangles stand for decisions you face (`choice nodes') and circles stand for events which might happen (`chance nodes').

Here is a more general explanation. Good assumes that rational agents maximize expected utility. So, if you are rational, then given some choice set $\mathcal{S}$, you will choose what seems best by your current lights: an action in $\mathcal{S}$ which maximizes expected utility with respect to your current credences. So the expected value of choosing now is the expected utility of one of the best actions in $\mathcal{S}$ relative to your current credences $p$:
\begin{equation*}
\max_{f \in \mathcal{S}} \mathbb{E}_{p} (f).
\end{equation*}
If, on the other hand, you learn that $E$ is the true event in our evidence partition, you update your credences $p$ to $\mathcal{P}_{E}$. Good assumes that in each state with non-zero probability, your updated credences are obtained from your current credences by conditionalizing, so $\mathcal{P}_{E} = p(\cdot \mid E)$. Good also assumes that learning is \text{cost-free}. This means that before and after learning, you choose among the same actions and outcomes have the same utilities. The only impact of the information is to change your credences.\footnote{Consider cases in which the information is not free (processing costs, library fees). In such cases, outcomes before and after learning do not have the same utility. \textcite[pp. 17-20]{Kadane2008} discuss this issue in detail. You might also ascribe negative utility to the information itself, for example because it makes you feel bad \parencite{Golman2017}. I set such cases aside and focus on the instrumental value of information.} 

After learning, you choose what seems best by your updated lights---an action in $\mathcal{S}$ which maximizes expected utility with respect to your updated credences:
\begin{equation*}
\max_{f \in \mathcal{S}} \mathbb{E}_{p(\cdot \mid E)} (f). 
\end{equation*}
You don't know yet which element of the evidence partition you will learn. But we can consider the \text{expected value} of acting after learning:
\begin{equation*}
\sum_{E \in \mathcal{E}} p(E) \max_{f \in \mathcal{S}} \mathbb{E}_{p(\cdot \mid E)} (f). 
\end{equation*}
Good completes the argument by proving that the expected value of acting after learning is always greater than or equal to the expected value of choosing now:
\begin{equation*}
\sum_{E \in \mathcal{E}} p(E) \max_{f \in \mathcal{S}} \mathbb{E}_{p(\cdot \mid E)} (f) \geq \max_{f \in \mathcal{S}} \mathbb{E}_{p} (f).
\end{equation*}
Moreover, this inequality is strict unless there is some action $f \in \mathcal{S}$ which is best regardless of which event in the evidence partition you learn---that is, unless the evidence partition is \text{irrelevant} for the choice set under consideration. So according to Good, the principle of maximizing expected utility entails:
\begin{quote}
\textbf{Value of Learning:}
\begin{enumerate}[label=\roman*., itemsep=0pt, topsep=0pt]

\item Rational agents are always permitted to accept free information before making a decision.

\item Rational agents are always required to accept free and relevant information before making a decision.

\end{enumerate}
\end{quote}

\subsection{What does the Argument show?}

Does Good's argument show that Value of Learning is correct? There are ways to push back. One might question the assumption that rational agents always maximize expected utility \parencite{Buchak2010, Campbell-Moore2020} or that rationality requires precise credences \parencite{Bradley2016, Wheeler2021}.\footnote{One could also question the assumption that learning can always be modeled as learning an element of a partition \parencite{Salow2019, Dorst2020, Das2023}, which is foreshadowed by \textcite[pp. 230-7]{Williamson2000}. Merely finitely additive probabilities can also lead to information aversion  \parencite{Kadane1996, Kadane2008}. Since I restrict attention to finite state spaces, the debate over finite versus countable additivity does not concern me here.} Relaxing these assumptions leads to cases where you can be required to reject free information. One might take this to question Value of Learning. However, one might also take this as a strike against alternatives to expected utility theory with precise credences. 

One might think that the permissibility of accepting free information before making a decision is independently plausible, a piece of common sense: `look before you leap'. It is a mark in favor of expected utility theory that it entails this piece of common sense and a problem for other decision theories if they conflict with it. From this perspective, Good's argument is not really an argument for Value of Learning but rather an argument for expected utility theory. This interpretation is suggested by \textcite{Kadane2008}:
\begin{quote}
So the question remains of whether it is reasonable to impose the requirement on a theory of rational decision making that it not require or permit paying not to see cost-free data. If it is, the only such theory known to us is Bayesian decision theory with a single countably-additive proper prior. \parencite[p. 33]{Kadane2008}
\end{quote}
Arguments along these lines are common. The general shape of the argument is that (a) Value of Learning is correct and (b) alternatives to expected utility theory are bad because they conflict with this. This assumes that (c) expected utility theory entails Value of Learning.\footnote{More examples: \textcite{Wakker1988} shows that violating the independence axiom of expected utility theory leads to situations in which agents reject free information and takes this to show that such violations are irrational. \textcite[p. 249]{AlNajjar2009} object to decision theories allowing for ambiguity aversion because they rationalize aversion to information ``which most economists would consider absurd or irrational''. \textcite{Briggs2015, Ahmed2016} object to risk-weighted expected utility (REU) theory because it leads to diachronic inconsistency and aversion to information. \textcite[187-9]{Buchak2013} also discusses how the value of information can be negative in REU theory and considers this to be a serious cost.}

These arguments are misguided. At least, they require serious qualification. This is because expected utility theory, supplemented with plausible assumptions, entails that there are cases in which we are rationally required to reject free information. Good's argument rests on the auxiliary assumption that \text{you are certain you will update by conditionalization}. This assumption should not be built into expected utility theory and there are good reasons to reject it. If we reject this assumption, expected utility maximizers with precise credences can be required to reject free and relevant information. So (c) is false: expected utility theory does not entail Value of Learning. I also argue that (a) is false: rational agents can be required to reject free information. So we should not take Value of Learning as axiomatic in our theories of instrumental rationality.

If you are already skeptical of Value of Learning, you might argue as follows: expected utility theory entails Value of Learning but Value of Learning is clearly false. There are many situations in real life where we are better off ignoring free information. Perhaps you think that the stranger at the horse track will try to deceive you. Therefore, we should reject expected utility theory and look for an alternative decision-theoretic framework, perhaps risk-weighted expected utility theory or imprecise credences. I agree that there are many situations in real life where we are better off ignoring free information. However, once we understand that expected utility theory does not entail Value of Learning, we can make sense of information aversion within the standard framework of expected utility theory and Bayesian epistemology.

\section{Against Good's Argument}

I explain why Good's argument presupposes that you are certain you will conditionalize (Immodesty). I argue that this assumption is implausible. Instead, we should assign some positive probability to not conditionalizing (Modesty). I show that expected utility maximization can require modest agents to reject free information.

\subsection{Good presupposes Immodesty}

As we have seen, Good's argument requires that whichever event in the evidence partition you learn, your future credences are obtained from your current credences by conditionalization: 

\begin{quote}
\textbf{The Equation:} $\mathcal{P}_{E} = p(\cdot \mid E)$ for every $E \in \mathcal{E}$.\footnote{Technically, Good's result requires only that this equality holds with probability one.}
\end{quote}

At first glance, one might think The Equation means that you are \text{actually} a conditionalizer. This is how the assumption is sometimes glossed in presentations of Good's argument.\footnote{For example, \textcite[p. 58]{Laffont1989} presents a result equivalent to Good's and writes that the agent under consideration ``revises his expectations by using Bayes's theorem". This sounds like we're assuming that the agent is actually a conditionalizer.} However, The Equation actually means that you assign subjective probability one to the event that you will conditionalize.\footnote{As \textcite[p. 247]{Skyrms1990} writes, ``the proof implicitly assumes not only that the decision maker is a Bayesian but also that he knows he will act as one. The decision maker believes with probability one that if he performs the experiment he will [...] update by conditionalization [...]''. \textcite[p. 283]{Huttegger2014} also makes this point.} In other words, \text{you are certain} you will conditionalize. This is because The Equation says that for \text{every} event you might learn, your new credences equal your old credences conditional on the learned event. Taken together, the events in the evidence partition sum to probability one. This means that in every state with positive probability, your new credences equal your old credences conditional on the true event in the evidence partition. Since states represent epistemic possibilities, you are certain you will conditionalize. 

So Good's argument presupposes
\begin{quote}
\textbf{Immodesty:} You are certain you will conditionalize.
\end{quote}

Here is another way to bring this out. You might in fact update by conditionalization. Nonetheless, you might  assign positive probability to a state in which you fail to conditionalize. In this case, The Equation does not hold and Good's argument does not go through. On the other hand, you might be certain you will conditionalize---and so satisfy The Equation---but fail to conditionalize in the actual world. This might be because you have assigned probability zero to an unforeseen failure of rationality. In this case, Good's argument still applies. What is at issue is not whether you will actually conditionalize but whether you are certain you will conditionalize. Even if you always conditionalize, you might have good reasons not to be certain of that.

\subsection{The Case for Modesty}

Immodesty is implausible. Instead, we should accept:
\begin{quote}
\textbf{Modesty:} There is some positive probability that you will not conditionalize. 
\end{quote}
To be clear, what I have in mind here is subjective probability, not objective chance. So to accept Modesty means to assign some positive credence to the possibility that you will not conditionalize.\footnote{Modesty has been defended before. For example, discussing whether we should defer to our future credences, \textcite[59-60]{Briggs2009} writes: ``Under all but the most ideal circumstances, agents will have reasons to suspect that future failures of conditionalization are in store''. \textcite{Pettigrew2020} points out that standard arguments for conditionalization assume `deterministic updating' and so leave no room for uncertainty about how you will update. \textcite{Lederman2015} draws on failures of common knowledge that we will conditionalize to construct counterexamples to Aumann's claim that rational agents cannot `agree to disagree'. \textcite{Cohen2020} discusses uncertainty about updating in the context of epistemic logic. \textcite[3]{Christensen2007} defends the broader claim that ``even an agent who is in fact cognitively perfect might, it would seem, be uncertain of this fact". Similar ideas are defended by many others, including \textcite{Carr2019, Bradley2020, Dorst2020}.} 

There are good reasons for Modesty. Moreover, these reasons flow from standard principles of Bayesian epistemology. Let me first be clear that it is by no means (physically or metaphysically) necessary that you will conditionalize. Rather, the claim that your new credences after learning are your old credences conditional on the learned event is a substantive claim about how your credences will evolve over time. The following passage by Ramsey makes the point clear:
\begin{quote}
[the degree of belief in p given q] is not the same as the degree to which [a subject] would believe p, if he believed q for certain; for knowledge of q might for psychological reasons profoundly alter his whole system of beliefs. \parencite[p. 21]{Ramsey1931}\footnote{\textcite{Diaconis1982} discuss this passage. Of course, the general idea is much more broadly recognized. For example, in \textit{The Portrait of a Lady}, Henry James writes about some piece of news: ``But it had been one thing to foresee it mentally, and it was another to behold it actually" \parencite[217]{James1882}.}
\end{quote}
So it is a consistent possibility that you fail to conditionalize. Many Bayesians are attracted to the principle of \text{regularity}, which says that you should assign positive prior probability to all consistent possibilities. This principle entails Modesty. 

More broadly, that you will conditionalize is an \text{empirical proposition}. Rationality should not require you to be certain that some empirical proposition is true. For example, if you suffer brain damage as the result of a stroke, you will likely not conditionalize. Plausibly, you should not be certain that you won't suffer brain damage in the future. Therefore, you should not be certain that you will conditionalize.\footnote{Note that the possibility of malfunction does not only apply to humans, but also to AI agents and plausibly to any kind of agent which is physically realized.}

We can make an even stronger case for Modesty. There is a long research tradition in psychology and cognitive science which aims to demonstrate that humans are not ideal Bayesian agents and so do not always conditionalize. There are a number of well-documented fallacies and heuristics which deviate from conditionalization. An example is the \text{base rate fallacy}, in which people ignore prior probabilities and so overestimate the probability of rare events \parencite{Kahneman1973}. Another example is the \text{gambler's fallacy}, which is when people think that a fair coin landing heads provides evidence that the next coin flip will land tails.

Once you learn about these empirical findings, it seems reasonable to believe that they might also apply to \text{yourself}. Therefore, you should assign some positive credence to not conditionalizing and accept Modesty. In addition to such general considerations, you might remember specific cases in which you did not conditionalize, but committed (say) the gambler's fallacy. If you have such evidence, this gives you another strong reason for Modesty. 

Perhaps you are quite confident of your future rationality. But even if you currently have no evidence that you might fail to conditionalize, it seems reasonable that you might obtain such evidence. For example, you might learn that you just took a drug which increases your susceptibility to the gambler's fallacy or that your brain is wired to misfire in certain situations.\footnote{The reason-impairing drug is inspired by \textcite{Christensen2007}.} Surely, if you learned something like this, you should decrease your credence that your future self will conditionalize. But Immodesty rules this out: once you assign probability zero to failures of conditionalization, then no matter what you learn, you will continue to assign probability zero to failures of conditionalization (if you actually conditionalize).\footnote{This is the key reason Bayesian epistemologists tend to be skeptical of assigning probability zero to any possible event. For example, \textcite[268]{Lewis1980} argues that regularity is ``required as a condition of reasonableness: one who started out with an irregular credence function (and who then learned from experience by conditionalizing) would stubbornly refuse to believe some propositions no matter what the evidence in their favor." }  This seems unreasonable---surely, there are some things you might learn that would make you doubt your own future rationality. Therefore, you should assign non-zero probability to failures of conditionalization, so Modesty follows.

The arguments above appeal to substantive constraints on prior probabilities. Subjective Bayesians reject such constraints beyond adherence to the axioms of probability. So subjective Bayesians will not be moved by my arguments. However, Immodesty is also a substantive constraint on prior probabilities and does \text{not} follow from the axioms of probability. Subjective Bayesians have no reason to accept this constraint.\footnote{As \textcite[315]{Hacking1967} points out, the axioms of probability don't entail that you will actually conditionalize, much less that you are certain of doing so: ``The idea of the model of learning is that $Prob(h/e)$ represents one's personal probability after one learns $e$. But formally, the conditional probability represents no such thing. [...] $Prob(h/e)$ stands merely for the quotient of two probabilities.''}

There are, of course, good reasons to think that rationality requires conditionalization, for  example diachronic coherence arguments \parencite{Lewis1999} and various accuracy arguments \parencite{Joyce1998, Greaves2005, Pettigrew2016}. Modesty is entirely consistent with this claim. Arguments for conditionalization aim to show:
\begin{quote}
\textbf{Conditionalization:} You should conditionalize.
\end{quote}
Good's argument relies on Immodesty, which says that you are certain you will conditionalize. Conditionalization does not entail Immodesty. We can accept that we should conditionalize but still have good reason to assign positive probability to failures of conditionalization in the future. This is because we might not be certain that our future selves will be rational. Indeed, as good Bayesians, we \text{should not} be certain that our future selves will be rational if our evidence suggests that we might not be.

Some philosophers have argued that the arguments for conditionalization only support the norm that you should \text{intend} or \text{plan} to conditionalize, rather than the norm that you should actually conditionalize.\footnote{This point is discussed, for example, by \textcite[pp. 187-88]{Pettigrew2016}.} If these philosophers are correct, then it is even harder to see any conflict between the arguments for conditionalization and Modesty. We can rationally plan to $\phi$ while also thinking that there is some positive probability that we will fail to $\phi$. For example, I can plan to run a 10K race while also thinking that there is some chance I won't make it to the end.\footnote{\textcite[p. 11-12]{Bratman1992} discusses similar examples and argues that one can plan to $\phi$ without believing that one will $\phi$.}

There are also reasons to doubt whether conditionalization is always rationally required. For example, \textcite{Douven2013} argues that an alternative to conditionalization, which he calls `Inference to the Best Explanation', leads you to converge to the truth faster in some circumstances. If you care about fast convergence, this might be a reason to use Douven's `Inference to the Best Explanation' instead of conditionalization. While this is no conclusive argument against conditionalization, it might perhaps instill some doubt about whether conditionalization is always rational. Plausibly, the right response to this normative uncertainty is to assign some positive probability to failures of conditionalization even if you are sure you will update rationally.

\subsection{Modesty entails Information Aversion}

Let us assume Modesty. I now explain how for modest agents, maximizing expected utility can require you to reject free information. The basic idea is quite simple. If you are modest, you assign some credence to the possibility that learning more information will lead you to make inferences which you do not currently endorse. This might lead you to make choices which, from your current point of view, seem like a bad idea. Therefore, you might be better off ignorant.

Suppose a fair coin will be flipped twice and Ann knows this. She chooses among bets on the second coin flip: a safe bet which always yields zero, a risky bet on heads and a risky bet on tails. Our choice set $\mathcal{S}$ is:
\begin{align*}
 & \texttt{safe}: \{ \$0 \textrm{ always} \}, \\
 & \texttt{risky-heads}: \{ \$1 \textrm{ if the second coin flip lands heads}, -\$2 \textrm{ otherwise} \}, \\
 & \texttt{risky-tails}: \{ \$1 \textrm{ if the second coin flip lands tails}, -\$2 \textrm{ otherwise} \}.
\end{align*}
Ann values money linearly and is an expected utility maximizer. She is also certain that her future self will be an expected utility maximizer.\footnote{I mean that she is certain she will pick one of the actions in $\mathcal{S}$ which maximizes expected utility relative to her updated credences---which might or might not be obtained from her current credences by conditionalization. The value Ann currently assigns to $f$ on the supposition of $E$ is the conditional expected utility $\mathbb{E}_{p(\cdot \mid E)}(f) = \sum_{\omega \in \Omega} p(\{\omega\} \mid E) f(\omega)$. \textcite{Gyenis2017} discuss conditional expectations in a much more general setting. The important point is that this conditional expected utility can come apart from the value Ann assigns to $f$ after actually learning $E$. I also assume that $\mathcal{S}$ does not include actions like `adopt credence $p$ after learning evidence $E$'. With such an extended option set, one can argue that certainty that one will maximize expected utility entails certainty that one will conditionalize \parencite{Brown1976}, although \textcite{Pettigrew2020} points out how uncertainty about updating complicates this argument. Thanks to an anonymous referee for pushing me to clarify how exactly I understand certainty that one maximizes expected utility.}

I offer Ann the following choice: She can either make her decision now or learn the outcome of the first coin flip and make her decision afterwards. If Ann makes her decision now, she will pick \texttt{safe}. So the expected value of choosing now is:
\begin{equation*}
\max_{f \in \mathcal{S}} \mathbb{E}_{p} (f) = \mathbb{E}_{p} (\texttt{safe}) = 0. 
\end{equation*}
What happens if Ann learns the outcome of the first flip and makes her decision afterwards? If Ann conditionalizes, she will choose the safe bet no matter what she learns since she regards the two coin flips as independent. So there is no reason for her to avoid learning. It can't help her, but it can't hurt her either.

But Ann is modest and assigns some positive probability to failures of conditionalization. In particular, Ann assigns some positive probability to committing the \text{gambler's fallacy}: after she learns that the first coin flip lands heads, she will become confident that the next coin flip will land tails and vice versa. In particular, Ann assigns some positive probability $\epsilon$ to the event that when she learns that the first coin flip lands heads, she will become .9 confident that the second coin flip will land tails and vice versa.

\begin{figure}[h]
\begin{tikzpicture}[grow=right, sloped]
\node[decision] {}
    child {
        node[chance] {}        
            child {
                node[chance] {}
                    child {
                    node[outcome, label=right: {Ann chooses \texttt{risky-heads}. EU: $-\frac{1}{2}$}] {}
                    edge from parent
                	   node[above] {Fallacy}
                    node[below]  {$\epsilon$}
                    }
                    child {
                    node[outcome, label=right: {Ann chooses \texttt{safe}. EU: 0}] {}
                    edge from parent
                	   node[above] {Bayes}
                    node[below]  {$1-\epsilon$}
                    }    
                edge from parent
                node[above] {Tails}
                node[below]  {$\frac{1}{2}$}
            }
            child {
                node[chance] {}
                    child {
                    node[outcome, label=right: {Ann chooses \texttt{risky-tails}. EU: $-\frac{1}{2}$}] {}
                    edge from parent
                	   node[above] {Fallacy}
                    node[below]  {$\epsilon$}
                    }
                    child {
                    node[outcome, label=right: {Ann chooses \texttt{safe}. EU: 0}] {}
                    edge from parent
                	   node[above] {Bayes}
                    node[below]  {$1-\epsilon$}
                    }
                edge from parent
                node[above] {Heads}
                node[below]  {$\frac{1}{2}$}
            }
            edge from parent 
            node[above] {Learn}
    }
    child {
        node[outcome, label=right: {Ann chooses \texttt{safe}. EU: 0}] {}    
        edge from parent         
            node[above] {Don't learn}
    };
\end{tikzpicture}
\caption{Ann's decision problem.}
\label{fig:Ann1}
\end{figure}
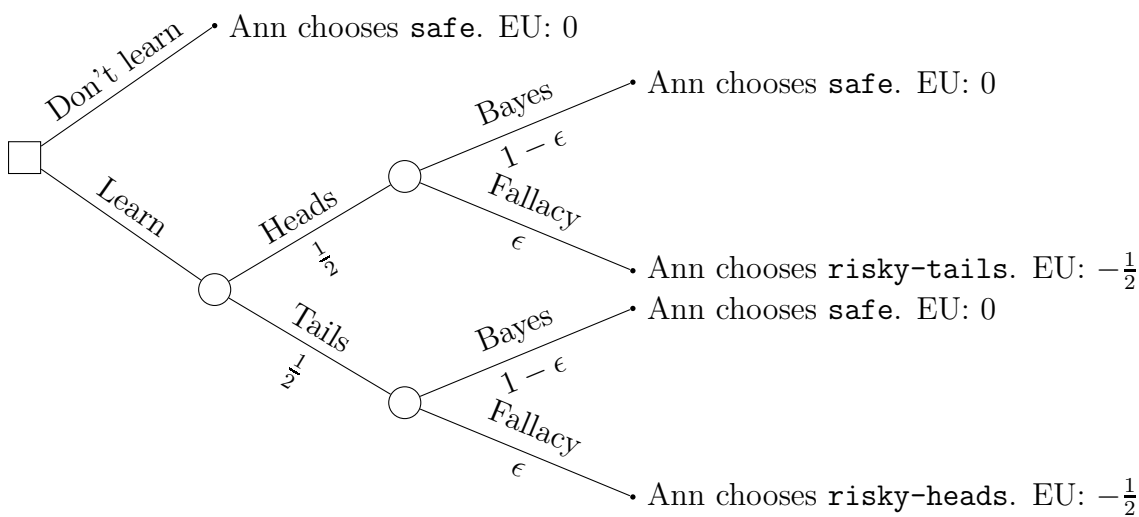

Now suppose Ann learns that the first coin flip lands heads and commits the gambler's fallacy. Relative to her updated credences, the risky bet on tails now looks like the best option. However, given Ann's current credences, the risky bet is the wrong choice. The situation is analogous if Ann learns that the first coin flip lands tails and commits the gambler's fallacy. Figure \ref{fig:Ann1} sums up Ann's situation.

\tikzstyle{level 1}=[level distance=2.5cm, sibling distance=3.5cm]
\tikzstyle{level 2}=[level distance=2.5cm, sibling distance=3cm]
\tikzstyle{level 3}=[level distance=3cm, sibling distance=2.5cm]

The expected value of learning and deciding afterwards is $-\frac{\epsilon}{2}$, \text{strictly worse} than the expected value of choosing now.\footnote{If Ann learns and decides afterwards, she chooses one of the risky options with probability $\epsilon$ and the safe option with probability $1 - \epsilon$. The risky options have expected utility $-\frac{1}{2}$ and the safe option has expected utility zero. So the expected value of learning and deciding afterwards is $\epsilon \times -\frac{1}{2} + (1 - \epsilon) \times 0 = -\frac{\epsilon}{2}$.} Learning the outcome of the first coin flip \text{can} hurt Ann but it can't help her, so she is better off ignorant. Since Ann can foresee all of this, it is rational for her to reject free information. So the principle of maximizing expected utility sometimes recommends rejecting free information.

When I say that the information is `free', I mean the same as Good: the information does not change the available actions or the utility function. The only impact of the information is to change Ann's credences. And it is not part of the example that Ann ever commits the gambler's fallacy. What makes it rational for Ann to reject the information is not that she actually deviates from conditionalization but that she assigns some positive probability to deviating from conditionalization.

I assume Ann is a `sophisticated chooser': she predicts her own future choices and takes these predictions into account when making her present decisions \parencite[35-6]{Hammond1988}.\footnote{\textcite[176]{Buchak2013} describes the debate around sophisticated choice in decision theory. In moral philosophy, there is a similar debate between actualism and possibilism about what you should do when your future self will act wrongly \parencite{Smith1976, Jackson1986}. \textcite{Louise2009, White2021} discuss the legitimate role of self-predictions in practical reasoning in more depth.} Since she predicts that her future self might be irrational, she has an incentive to prevent her future self from making bad choices. So Ann faces a predicament similar to Odysseus in Greek mythology. She predicts that learning might compromise her future rationality, so she is better off ignorant. 

You might complain that this example is a bit weird. Since Ann regards the two coin flips as independent, there is no way that learning the outcome of the first coin flip could help her make a better choice. At best, the information is neutral. In other words, if Ann is certain she will conditionalize, learning the outcome of the first coin flip is not \text{relevant} to her choice set. However, we can modify the example so that the information is relevant to her choice set but the principle of maximizing expected utility still recommends rejecting the information.

Again, a coin will be flipped twice and Ann must decide among several bets on the second coin flip. There is a safe bet, a slightly risky bet on heads, a slightly risky bet on tails, a very risky bet on heads and a very risky bet on tails in our choice set $\mathcal{S}$:
\begin{align*}
 & \texttt{safe}: \{ \$0 \textrm{ always} \}, \\
 & \texttt{heads}: \{ \$1 \textrm{ if the second coin flip lands heads}, -\$1 \textrm{ otherwise} \}, \\
 & \texttt{tails}: \{ \$1 \textrm{ if the second coin flip lands tails}, -\$1 \textrm{ otherwise} \}, \\
  & \texttt{v-risky-heads}: \{ \$2 \textrm{ if the second coin flip lands heads}, -\$10 \textrm{ otherwise} \}, \\
 & \texttt{v-risky-tails}: \{ \$2 \textrm{ if the second coin flip lands tails}, -\$10 \textrm{ otherwise} \}.
\end{align*}

This time, Ann does not consider the coin to be fair but thinks that the coin has an unknown bias. The coin might be fair, it might be biased towards heads or it might be biased towards tails---she has no idea. Again, I offer Ann the following choice: She can either make her decision now or learn the outcome of the first coin flip and make her decision afterwards. 

Since the coin has an unknown bias, observing the outcome of the first coin flip is informative for Ann. In particular, let us assume that, conditional on the first coin flip landing heads, Ann thinks that the second coin flip lands heads with probability $\frac{2}{3}$. The same goes for tails.\footnote{These probabilities can be derived from the `rule of succession' \parencite{Zabell1989}.}

If Ann makes her decision now, she is indifferent between $\texttt{safe}$, $\texttt{heads}$ and $\texttt{tails}$. So the expected value of choosing now is:
\begin{equation*}
\max_{f \in \mathcal{S}} \mathbb{E}_{p} (f) = \mathbb{E}_{p} (\texttt{safe}) = 0. 
\end{equation*}
If Ann observes the outcome of the first coin flip, things are more interesting. Suppose Ann will conditionalize. Then if the coin lands heads, Ann will think that the coin is probably biased towards heads, so the slightly risky bet on heads will seem best to her. The very risky bet on heads will still seem too risky. The situation is analogous if the coin lands tails. 

But Ann is modest and assigns some positive probability $\epsilon$ to the event that she \text{overweights} the evidence: when she learns that the first coin flip lands heads, she becomes .9 confident that the second coin flip will land heads and vice versa. So Ann takes the evidence into account, but thinks that she might be overconfident in how she does it. There are several reasons for why Ann might do this. She might commit some version of the \text{base rate fallacy}, ignoring or underweighting prior probabilities. Or she might be susceptible to some form of the \text{hot hand fallacy}, believing that `streaks' of successive heads are more likely than warranted by her evidence.

\begin{figure}[h]
\begin{tikzpicture}[grow=right, sloped]
\node[decision] {}
    child {
        node[chance] {}        
            child {
                node[chance] {}
                    child {
                    node[outcome, label=right: {Ann chooses \texttt{v-risky-tails}. EU: $-2$}] {}
                    edge from parent
                	   node[above] {Fallacy}
                    node[below]  {$\epsilon$}
                    }
                    child {
                    node[outcome, label=right: {Ann chooses \texttt{tails}. EU: $\frac{1}{3}$}] {}
                    edge from parent
                	   node[above] {Bayes}
                    node[below]  {$1-\epsilon$}
                    }    
                edge from parent
                node[above] {Tails}
                node[below]  {$\frac{1}{2}$}
            }
            child {
                node[chance] {}
                    child {
                    node[outcome, label=right: {Ann chooses \texttt{v-risky-heads}. EU: $-2$}] {}
                    edge from parent
                	   node[above] {Fallacy}
                    node[below]  {$\epsilon$}
                    }
                    child {
                    node[outcome, label=right: {Ann chooses \texttt{heads}. EU: $\frac{1}{3}$}] {}
                    edge from parent
                	   node[above] {Bayes}
                    node[below]  {$1-\epsilon$}
                    }
                edge from parent
                node[above] {Heads}
                node[below]  {$\frac{1}{2}$}
            }
            edge from parent 
            node[above] {Learn}
    }
    child {
        node[outcome, label=right: {Ann chooses \texttt{safe}. EU: 0}] {}    
        edge from parent         
            node[above] {Don't learn}
    };
\end{tikzpicture}
\caption{Ann's other decision problem.}
\label{fig:Ann2}
\end{figure}
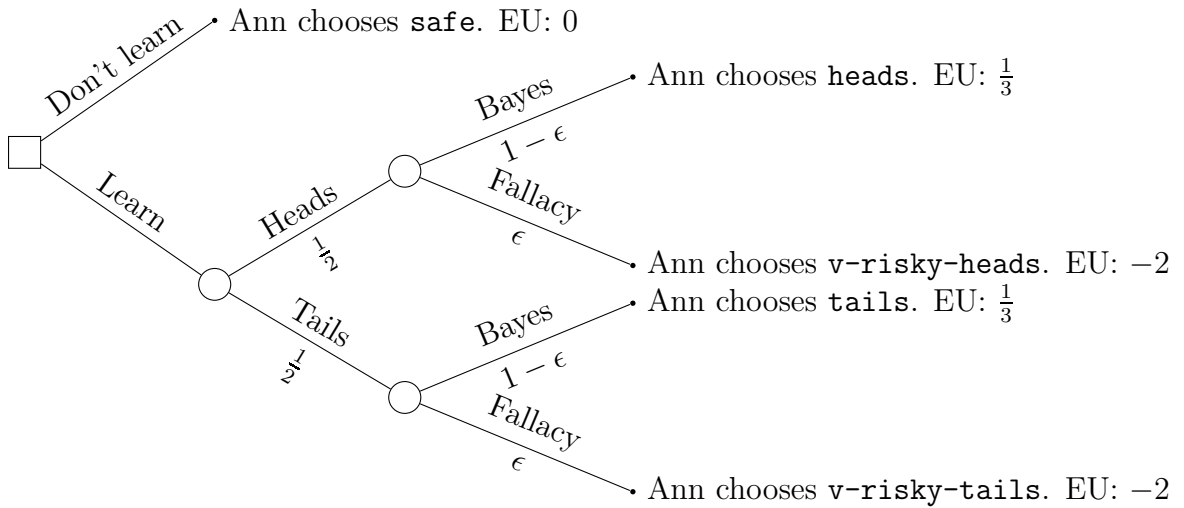

Suppose Ann observes the first coin flip landing heads. If she conditionalizes, she will take the slightly risky bet on heads. But if she is overconfident, she will choose the very risky bet on heads, which looks like a bad choice from her current point of view. A similar story applies if Ann observes the first coin flip landing tails. Figure \ref{fig:Ann2} sums up Ann's situation.

The expected value of learning is $\frac{1}{3}(1 - \epsilon) - 2 \epsilon$.\footnote{If Ann learns and decides afterwards, she chooses one of the very risky options (\texttt{v-risky-heads},  \texttt{v-risky-tails}) with probability $\epsilon$ and one of the less risky options (\texttt{heads}, \texttt{tails}) with probability $1 - \epsilon$. The very risky options have expected utility $-2$ and the less risky options have expected utility $\frac{1}{3}$. So the expected value of learning and deciding afterwards is $\epsilon \times - 2 + (1 - \epsilon) \times \frac{1}{3} = \frac{1}{3}(1 - \epsilon) - 2 \epsilon$.} So if Ann thinks the probability of overconfidence is more than $\frac{1}{7}$, the expected value of learning and making her decision afterwards is worse than the expected value of deciding now.\footnote{Choosing now has expected value zero and $0 > \frac{1}{3}(1 - \epsilon) - 2 \epsilon \iff \epsilon > \frac{1}{7}$.} So even if learning could be informative, expected utility maximizers can be required to reject free information. Again, it is not part of the example that Ann actually overweights her evidence but only that she assigns some probability to doing so.

These example show two things. First, the principle of maximizing expected utility does not imply Value of Learning. Expected utility maximizers are not always permitted to accept free and relevant information. Good's argument essentially depends on the assumption of Immodesty. If we assume Modesty, the principle of maximizing expected utility can require agents to reject free and relevant information.

Second, Value of Learning is false: Ann is rational but not permitted to accept free and relevant information. This might seem contentious. Whether or not rationality requires us to always conditionalize, the gambler's fallacy certainly looks irrational. So Ann thinks there is some probability that her future self will be irrational. However, the fact that Ann thinks her future self might be irrational does not entail that Ann is currently irrational. Rationality does not require you to be certain that your future self will be rational.

We can suppose that Ann has good evidence she might commit the gambler's fallacy. All her friends have committed it and she thinks they are relevantly similar to her. In this situation, it is implausible to think that Ann must be certain that her future self will conditionalize. Rather, if she is a good Bayesian, she should take her evidence into account and be modest. We can also suppose that Ann plans to conditionalize. Furthermore, we can suppose that in the actual word, Ann always manages to follow her plan. She just thinks that there is some chance she might fail. This does not seem like a failure of rationality. Therefore, we should let Value of Learning go even if we accept expected utility theory with precise credences and information which partitions logical space. 

\subsection{Information Aversion in the Real World}

Moreover, we can use Modesty to make sense of real-world cases of information aversion. I will briefly illustrate this with medical testing, blind grading, checking one's stock portfolio and resisting manipulation.

People sometimes reject medical tests. There are several reasons: mistrust of doctors, fear of bad news and so on \parencite[p. 365]{Hertwig2016}.\footnote{Information aversion with respect to medical tests is discussed by \textcite{Osimani2012}, \textcite{Jouini2018} and many others.}  Modesty suggests another reason. People could fear that the test results might lead them (or their doctors) to draw  inferences which they do not currently endorse, resulting in unnecessary treatment and further testing. For example, imagine you learn that a certain marker has increased in your blood test since last time but is still in the normal range. Learning this information might lead you to suspect a worrying trend where there are only random fluctuations. As a result, you might want another test soon which is unnecessary.
 
Blind grading is often considered good practice. Why is it bad to know the student's names? The standard explanation is that blind grading reduces bias. For example, I might give too much weight to the fact that George got an `A' on the first paper and treat it as better evidence than it is that his current paper deserves a good grade.

It is sometimes suggested that you shouldn't check your stock portfolio daily. One reason is that it might stress you. However, another reason is that you might be tempted to change the allocation of your portfolio in a way that you currently view as a bad idea. This, in turn, can be explained by the risk of overweighting the significance of small fluctuations.\footnote{In a best-selling popular science book on computer science and decision theory, \textcite[148]{Christian2016} write: ``If you want to be a good intuitive Bayesian---if you want to naturally make good predictions, without having to think about what kind of prediction rule is appropriate---you need to protect your priors. Counterintuitively, that might mean turning off the news''. They do not consider how we can make sense of this idea without contradicting Good's theorem. Modesty offers an elegant way of doing so.}

It appears rational to refuse information designed to manipulate you. For example, one reason to avoid social media is that the information shown on your feed is designed to influence your behavior: to make you buy the products advertised there, to make you spend even more time on the platform and so on \parencite[69-76]{Veliz2020}. Even if we sidestep issues of misinformation and assume the information you see on your feed is accurate, the fact that this information is designed to influence your behavior by companies which do not have your best interest at heart is a reason to stop looking. A similar case is when you refuse to talk to a manipulative person. Even if everything the manipulative person says is true, you might be better off not listening. This is because you might suspect that you will not update rationally on information designed to manipulate you.

Many other examples of information aversion in real life can be explained along similar lines.\footnote{For example elite-group ignorance, which \textcite{Kinney2021} explain using risk-weighted expected utility theory. \textcite{YongForth} critically discusses this explanation.} There are, of course, competing explanations: perhaps people deviate from expected utility theory, have imprecise credences or assign negative utility to bad news. But in the examples above, it seems independently plausible that we assign some probability to \text{overweighting evidence}: giving too much weight to the result of medical tests, the past performance of our students, small fluctuations in our stock portfolio and the information on our feed. When we reflect on how much weight we should assign to this information, we might conclude: `a little bit, but not very much'. But once we actually learn the information, we might assign more weight to it than we have previously considered rational. Imagine a positive result on a medical test slightly increases your probability of serious illness. Before doing the test, you might calmly assign conditional probabilities which reflect this slight increase. But when you learn that the test actually turned out positive, you might increase our probability of serious illness more than you previously considered warranted.

If we model overweighting evidence as deviation from conditionalization, we have seen that it can be a good idea to reject free information. So Modesty can explain these examples of information aversion in the real world in a way that seems to get at the heart of the matter. On the other hand, explaining these cases by saying, for example, that people are not expected utility maximizers seems to have less independent motivation. So while I have no conclusive argument that Modesty is the correct explanation for these cases, it seems like a particularly plausible candidate.\footnote{Thanks to an anonymous referee for pushing me to clarify why Modesty is a plausible explanation of these cases of information aversion.}      

\section{Value of Information Generalized}

Surely, modest agents are not always required to reject free information. Even if you have some uncertainty about how you will update, this does not mean that you are always better off ignorant. But how should modest agents decide when to learn more information? And how general is the link between information aversion and Modesty? I answer these questions by generalizing the value of information to modest agents.

\subsection{Good's Value of Information}

Good's argument gives us a way to measure the value of information. To state this idea, it is useful to introduce an additional bit of notation. I write $\mathcal{P}(\cdot \mid \mathcal{E})$ for your credences updated by conditionalization on the evidence partition $ \mathcal{E}$. This is a random variable which takes different probability functions as values in different state.\footnote{More rigorously, we can define $\mathcal{P}(\cdot \mid \mathcal{E})$ as the random variable which maps each $\omega \in \Omega$ to $p(\cdot \mid E)$ for the unique $E \in \mathcal{E}$ such that $\omega \in E$.} Then, we can define the value of information as follows \parencite{Blackwell1951, Raiffa1961, Howard1966}:\footnote{\textcite{LeCam1996} sketches the history of this concept, which apparently goes back to an unpublished RAND memorandum entitled `Reconnaissance in Game Theory' based on suggestions by von Neumann \parencite{Bohnenblust1949}.}
\begin{definition}\label{goodvoi}
The value of information for $\mathcal{E}$ is:
\begin{equation*}
Val_{Good}(\mathcal{E}) = \mathbb{E}_p \left( \max_{f \in \mathcal{S}} \mathbb{E}_{\mathcal{P}(\cdot \mid \mathcal{E})} (f) \right)
 - \max_{f \in \mathcal{S}} \mathbb{E}_{p} (f).
\end{equation*}
\end{definition}
This is the difference between the expected value of choosing after learning and the expected value of choosing now. It measures how much you expect the information to improve your decision. In this context, Value of Learning is captured by the fact that for any evidence partition $\mathcal{E}$, $Val_{Good}(\mathcal{E}) \geq 0$. In slogan form: `the value of information is always non-negative'.

This concept is useful because it allows us to say \text{how much} you should value learning the answer to a question. It also allows us to compare the value of learning the answers to different questions. This means we can formalize tradeoffs between acting now versus learning more and acting later even if learning is costly. Such tradeoffs are ubiquitous. In many real-life contexts, such as drug trials, we have to decide how much to sacrifice for learning more information.\footnote{Such decision problems can be formalized as `multi-armed bandits' in which one must balance \text{exploiting}, acting according to one's current best estimate, and \text{exploring} new and potentially better options \parencite{Lattimore2020}. The value of information can be used to define optimal solutions to such problems.} So it is not surprising that the value of information is widely used in economics and artificial intelligence.\footnote{\textcite[pp. 628-33]{Russell2018} discuss the value of information in AI research. \textcite{Hadfield2017} discuss a model of how to ensure that AI agents always defer to humans which  relies on the value of information being non-negative. In addition, the value of information is relevant to much other work, for example in the philosophy of language \parencite{Van2003}, to discussions about `longtermism' in ethics \parencite{AskellForth} and the epistemology of disagreement \parencite{DorstForthcoming}.} However, the standard way of defining the value of information presupposes Immodesty. 

\subsection{General Value of Information}

Here is a proposal for how we can define the value of information in a more general way. I write  $\mathcal{P}_{\mathcal{E}}$ for your credences updated on the evidence partition $\mathcal{E}$ without assuming you are certain you will update by conditionalization. This is a random variable whose values are different probability functions in different states.\footnote{On each $E \in \mathcal{E}$, $\mathcal{P}_{\mathcal{E}}$ agrees with $\mathcal{P}_{E}$ as defined in section (\ref{terminology}).}

Suppose you will learn the true element of some evidence partition $\mathcal{E}$. Then you update your credences in some way---perhaps you conditionalize, perhaps you do something different---and choose the action which maximizes expected utility relative to your updated credences:
\begin{equation*}
\argmax_{f \in \mathcal{S}} \mathbb{E}_{\mathcal{P}_{\mathcal{E}}} (f).\footnote{The term $\argmax_{x \in X} g(x)$ denotes the \text{argument of the maximum}: the $x \in X$ such that $g(x)$ is maximal.}
\end{equation*}
This expression will usually denote different actions in different states, because you might learn different events and update on those events in different ways. I assume that there is a unique best action in every state.\footnote{So for every $\omega \in \Omega$, there is a unique $f^* \in \mathcal{S}$ maximizing $\mathbb{E}_{\mathcal{P}_{\mathcal{E}}(\omega)} (\cdot)$. Recall that $\mathcal{P}_{\mathcal{E}}$ is a function from states to probability functions, so $\mathcal{P}_{\mathcal{E}}(\omega)$ is a particular probability function---the credence you adopt after learning the true element of $\mathcal{E}$ in state $\omega$. I do not consider cases where two actions are tied for the best action because in such cases, we would need to consider how to break the tie (introduce a selection function), which leads to additional complications. \textcite[190]{Buchak2013} provides relevant discussion and references on how indifference complicates sophisticated choice.}
 
We are interested in evaluating how good this action is from your current perspective, so we consider the expected utility of this action given your current credences:
\begin{equation*}
 \mathbb{E}_{p} \left( \argmax_{f \in \mathcal{S}} \mathbb{E}_{\mathcal{P}_{\mathcal{E}}} (f) \right).
\end{equation*}
This is the expected utility of the action you think you will actually do after learning. We model a `sophisticated chooser': our agent predicts her future choices and takes this information into account when making present decisions. I propose the following definition: 
\begin{definition}\label{generalvoi}
The \text{general value of information} for $\mathcal{E}$ is:
\begin{equation*}
Val_{General}(\mathcal{E}) =  \mathbb{E}_{p} \left( \argmax_{f \in \mathcal{S}} \mathbb{E}_{\mathcal{P}_{\mathcal{E}}} (f) \right)
 - \max_{f \in \mathcal{S}} \mathbb{E}_{p} (f).
\end{equation*}
\end{definition}

This measures the difference between the expected utility of your current best action and the expected utility of the action that you think you will choose after learning. In contrast to Good, I do not assume that you are certain you will conditionalize. I still assume you are certain you will maximize expected utility.

If we assume Immodesty, my proposal is identical to Good's: 
\begin{theorem}
If $\mathcal{P}_{E} = p( \cdot \mid E)$ for all $E \in \mathcal{E}$, then $Val_{General}(\mathcal{E}) = Val_{Good}(\mathcal{E})$.
\end{theorem}
In contrast to Good's value of information, the general value of information can be negative. We have seen this in the examples above. But it is not always negative. This is shown by the second example above. When Ann considers the possibility of overconfidence sufficiently unlikely, she is better off observing the first coin flip. In (slightly clunky) slogan form: `the value of information is sometimes negative, but not always. It depends'.

We can also say something about how general the link between Modesty and  information aversion is. For this purpose, I make two additional assumptions. \text{Utility Richness} says that for every $x \in [0,1]$, there is some outcome $o \in \mathscr{O}$ such that $u(o) = x$.\footnote{Our outcome space could contain lotteries which yield outcome $b$ with probability $p$ and outcome $w$ with probability $(1-p)$. In this case, we only need two `primitive' outcomes $b$ and $w$ with $u(b) > u(w)$ to obtain rich utilities.} \text{Evidential Independence} says that conditional on the learned event, your updated credences are independent of what action is best.\footnote{More precisely, for every $E \in \mathcal{E}$, your updated credences $\mathcal{P}_{E}$, which determine which action you will choose after learning, are independent of all $f \in \mathcal{S}$ conditional on $E$. This means, in particular, that for all $f,g \in S$, $\mathbb{E}_{p(\cdot \mid E \cap \textrm{choose $f$})} (g) = \mathbb{E}_{p(\cdot \mid E)} (g)$, where `choose $f$' is the event that you choose action $f$ after learning. The intuition is that learning that you choose a particular action after learning $E$ does not affect the expected utility of actions beyond learning $E$. It would be interesting to investigate cases where deviations from conditionalization are systematically correlated with which actions are best. In this paper, I focus on the simple case where Evidential Independence holds.} Evidential Independence rules out cases where you deviate from conditionalizing because you become more certain of the truth. For example, you might observe that a fair coin lands heads and be able to foresee that it lands tails next. If we agree that your evidence is that the coin lands heads on the first flip, you do not update by conditionalizing on your evidence. However, clairvoyance can lead you to make better decisions than conditionalization. In contrast, I consider deviations from conditionalization which are not systematically correlated with which action is actually best. I have implicitly made this assumption earlier: Ann is equally likely to commit the gambler's fallacy whether the second coin flip lands heads or tails.

Assuming Evidential Independence, we can write $Val_{General}(\mathcal{E})$ as:
\begin{equation*}
Val_{General}(\mathcal{E}) =  \sum_{E \in \mathcal{E}} p(E) \sum_{i=1}^n p(\textrm{choose $f_i$} \mid E) \mathbb{E}_{p(\cdot \mid E)} (f_i) - \max_{f \in \mathcal{S}} \mathbb{E}_{p} (f),
\end{equation*}
where `$\textrm{choose $f_i$}$' denotes the event that you choose action $f_i$ after learning $E$, which means that $f_i$ maximizes expected utility relative to your updated credences after learning $E$. Note that $p(\textrm{choose $f_i$} \mid E)$ is your current conditional probability that you will choose action $f_i$ after learning $E$. You evaluate how good this action is by its conditional expected utility $\mathbb{E}_{p(\cdot \mid E)} (f_i)$ given your current credences. If you do not conditionalize, this conditional expected utility can come apart from the unconditional expected utility of the action according to your updated credences.\footnote{Thanks to an anonymous referee for suggesting to make the formula for $Val_{General}(\mathcal{E})$ more explicit and see Lemma (\ref{lemma1}).}

We can show the following:
\begin{theorem}
Assuming Utility Richness and Evidential Independence, for every modest agent, there is some choice set where $Val_{General}(\mathcal{E}) < 0$.
\end{theorem}
\noindent Given our assumptions, any positive probability of not conditionalizing leads to information aversion. It does not matter why we are modest, as long as Evidential Independence holds. The examples above demonstrated how \text{psychological uncertainty} about whether you will update rationally leads to information aversion. The theorem shows that even if you are certain that your future self will be rational, \text{normative uncertainty} about whether conditionalization is rational leads to information aversion. So we have a very general argument against Value of Learning. This also means that we cannot rescue Good's argument by saying that while we might not be \text{certain} we will conditionalize, we are \text{very confident} we will conditionalize. (Perhaps we have a `default entitlement' to believe in our future rationality.) Any non-zero probability of failing to conditionalize means trouble for Good.

We can also show:
\begin{theorem}
Assuming Evidential Independence, $\mathcal{E}$, $Val_{General}(\mathcal{E}) \leq Val_{Good}(\mathcal{E})$ for every evidence partition $\mathcal{E}$.
\end{theorem}
\noindent Doubts about how you will update cannot increase the value of information. Note that one might take this theorem as a reason to think that you \text{should} conditionalize, at least relative to the assumption of Evidential Independence. But we are often not sure whether we will be rational in the future and cannot change anything about that. If we are in such a situation, the theorem tells us that we should value learning less than if we were certain that we would conditionalize.

\section{Conclusion}

Good argues that the principle of maximizing expected utility entails Value of Learning: rational agents are always permitted to accept free information and required to accept information which is free and relevant. I have argued that the principle of maximizing expected utility does not entail Value of Learning and that Value of Learning is false. The key observation is that Good's argument only works if we are certain that we will update by conditionalization but  we have good reason not to be.

What follows? First, we can give better advice to modest agents: sometimes, they are better off ignorant. Since we arguably are---and should be---modest, this advice applies to us. Sometimes, we are better off ignorant. Sometimes, we should avert our eyes and stuff our ears with wax to avoid learning the song of the Sirens. Second, proponents of expected utility theory should be careful when objecting to alternative decision-theoretic frameworks on the grounds that these frameworks sometimes permit or require agents to avoid free information. Properly understood, expected utility theory does the same. So this objection loses much of its dialectical force. Third,  information aversion is a feature and not a bug. Plausible arguments from Bayesian epistemology push us towards Modesty. And once we accept Modesty, we can explain many instances of information aversion in the real world which would otherwise be puzzling. By going beyond Good, we end up with a better decision theory.

\section*{Appendix}

\begin{theorema}
If $\mathcal{P}_{E} = p( \cdot \mid E)$ for all $E \in \mathcal{E}$, then $Val_{General}(\mathcal{E}) = Val_{Good}(\mathcal{E})$.
\end{theorema}

\begin{proof}
It suffices to show that if $\mathcal{P}_{E} = p( \cdot \mid E)$ for all $E \in \mathcal{E}$, then
\begin{equation}\label{voiequal}
\mathbb{E}_p \left( \argmax_{f \in \mathcal{S}} \mathbb{E}_{\mathcal{P}_{\mathcal{E}}} (f) \right) = \mathbb{E}_{p} \left( \max_{f \in \mathcal{S}} \mathbb{E}_{\mathcal{P}(\cdot \mid \mathcal{E})} (f) \right).
\end{equation}
By the law of total expectation, since $\mathcal{E}$ is a partition with $p(E) > 0$ for all $E \in \mathcal{E}$,\footnote{In general, the law of total expectation says that for any random variables $X$ and $Y$, $\mathbb{E}(X) = \mathbb{E}(\mathbb{E}(X \mid Y))$ \parencite[403]{Pitman1993}. I use the special case where $\mathcal{E}$ is a partition with $p(E) > 0$ for all $E \in \mathcal{E}$ and $X$ a random variable. Then $\mathbb{E}(X) = \sum_{E \in \mathcal{E}} p(E) \mathbb{E}(X \mid E)$. On every $E \in \mathcal{E}$, $\argmax_{f \in \mathcal{S}} \mathbb{E}_{\mathcal{P}_{\mathcal{E}}}$ agrees with $\argmax_{f \in \mathcal{S}} \mathbb{E}_{\mathcal{P}_{E}}$.}
\begin{equation*}
 \mathbb{E}_{p} \left( \argmax_{f \in \mathcal{S}} \mathbb{E}_{\mathcal{P}_{\mathcal{E}}} (f) \right) = \sum_{E \in \mathcal{E}} p(E) \mathbb{E}_{p(\cdot \mid E)} \left( \argmax_{f \in \mathcal{S}} \mathbb{E}_{\mathcal{P}_{E}} (f)  \right).
  \end{equation*}
We assume that $\mathcal{P}_{E} = p(\cdot \mid E)$ for all $E \in \mathcal{E}$, so
\begin{equation*}
 \sum_{E \in \mathcal{E}} p(E) \mathbb{E}_{p(\cdot \mid E)} \left( \argmax_{f \in \mathcal{S}} \mathbb{E}_{\mathcal{P}_{E}} (f) \right) =  \sum_{E \in \mathcal{E}} p(E) \mathbb{E}_{p(\cdot \mid E)} \left( \argmax_{f \in \mathcal{S}} \mathbb{E}_{p(\cdot \mid E)} (f) \right) .
\end{equation*}
Now $\mathbb{E}_{p(\cdot \mid E)} \left( \argmax_{f \in \mathcal{S}} \mathbb{E}_{p(\cdot \mid E)} (f) \right) = \max_{f \in \mathcal{S}}  \mathbb{E}_{p(\cdot \mid E)} (f)$, so
\begin{equation*}
 \sum_{E \in \mathcal{E}} p(E) \mathbb{E}_{p(\cdot \mid E)} \left( \argmax_{f \in \mathcal{S}} \mathbb{E}_{p(\cdot \mid E)} (f) \right) =  \sum_{E \in \mathcal{E}} p(E) \max_{f \in \mathcal{S}}  \mathbb{E}_{p(\cdot \mid E)} (f)  = \mathbb{E}_{p} \left( \max_{f \in \mathcal{S}} \mathbb{E}_{\mathcal{P}(\cdot \mid \mathcal{E})} (f) \right),
\end{equation*}
which shows that (\ref{voiequal}) holds.
\end{proof}

\begin{lemma}\label{lemma1}
Assuming \text{Evidential Independence}, for every $f \in \mathcal{S}$,
\begin{equation*}
\mathbb{E}_{p(\cdot \mid E)} \left( \argmax_{f \in \mathcal{S}} \mathbb{E}_{\mathcal{P}_{E}} (f) \right) = \sum_{i=1}^n p(\textrm{choose $f_i$} \mid E) \mathbb{E}_{p(\cdot \mid E)} (f_i),
\end{equation*}
where  `$\textrm{choose $f_i$}$' is the event that you choose action $f_i$ after learning $E$.
\end{lemma}

\begin{proof}
Suppose the range of $\argmax_{f \in \mathcal{S}} \mathbb{E}_{\mathcal{P}_{\mathcal{E}}}$ is $f_1, ... , f_n$. Intuitively, these are the actions you might choose after learning. Let us abbreviate the event $\argmax_{f \in \mathcal{S}} \mathbb{E}_{\mathcal{P}_{\mathcal{E}}} = f_i$ by `$\textrm{choose $f_i$}$'. Intuitively, this is the event that you choose action $f_i$ after learning $E$. (Recall that there is always a unique best action after learning.)

\text{Evidential independence} holds if for every $E \in \mathcal{E}$, $\mathcal{P}_{E}$ is independent of all $f \in \mathcal{S}$ conditional on $E$.\footnote{\textcite[400]{Pitman1993} defines conditional independence for random variables.} This means, in particular, that for all $f \in S$ and $f_i$ with $1 \leq i \leq n$, $\mathbb{E}_{p(\cdot \mid E \cap \textrm{choose $f_i$})} (f) = \mathbb{E}_{p(\cdot \mid E)} (f)$. The intuition is that relative to your prior, the event that you choose a particular action after learning $E$ does not affect the expected utility of actions beyond learning $E$.

We want to show:
\begin{equation}\label{lemma1goal}
\mathbb{E}_{p(\cdot \mid E)} \left( \argmax_{f \in \mathcal{S}} \mathbb{E}_{\mathcal{P}_{E}} (f) \right) = \sum_{i=1}^n p(\textrm{choose $f_i$} \mid E) \mathbb{E}_{p(\cdot \mid E)} (f_i).
\end{equation}

Since the events $\textrm{`choose $f_1$'}, .... , \textrm{`choose $f_n$'}$ form a partition, we can apply the law of total expectation:
\begin{equation*}
\mathbb{E}_{p(\cdot \mid E)} \left( \argmax_{f \in \mathcal{S}} \mathbb{E}_{\mathcal{P}_{E}} (f) \right) = \sum_{i=1}^n p(\textrm{choose $f_i$} \mid E) \mathbb{E}_{p(\cdot \mid E \cap \textrm{choose $f_i$})} \left( \argmax_{f \in \mathcal{S}} \mathbb{E}_{\mathcal{P}_{E}} (f) \right).
\end{equation*}
Now $\mathbb{E}_{p(\cdot \mid E \cap \textrm{choose $f_i$})} \left( \argmax_{f \in \mathcal{S}} \mathbb{E}_{\mathcal{P}_{E}} (f) \right) = \mathbb{E}_{p(\cdot \mid E \cap \textrm{choose $f_i$})} \left( f_i \right)$ by the definition of $\textrm{`choose $f_i$'}$, so
\begin{equation*}
\mathbb{E}_{p(\cdot \mid E)} \left( \argmax_{f \in \mathcal{S}} \mathbb{E}_{\mathcal{P}_{E}} (f) \right) = 
\sum_{i=1}^n p(\textrm{choose $f_i$} \mid E) \mathbb{E}_{p(\cdot \mid E \cap \textrm{choose $f_i$})} (f_i).
\end{equation*}
By Evidential Independence, $\mathbb{E}_{p(\cdot \mid E \cap \textrm{choose $f_i$})} (f_i) = \mathbb{E}_{p(\cdot \mid E)} (f_i)$, so (\ref{lemma1goal}) holds.
\end{proof}

\begin{theorema}
Assuming Utility Richness and Evidential Independence, for every modest agent, there is some choice set where $Val_{General}(\mathcal{E}) < 0$.
\end{theorema}

\begin{proof}
An agent is \text{modest} iff she assigns some positive probability to not conditionalizing. So for some evidence partition $\mathcal{E}$, there is some $E \in \mathcal{E}$ such that with positive probability, $\mathcal{P}_{E} \not = p( \cdot \mid E)$. This means that for some $\omega \in E$ with $p(\omega) > 0$, $\mathcal{P}_{E}(\omega)(A) \not = p(A \mid E)$ for some event $A$. I write $p_{E}$ for $\mathcal{P}_{E}(\omega)$ and assume $p_{E}(A) > p(A \mid E)$. (In the other case, the proof is analogous.)

We want to show that there is a choice set where $Val_{General}(\mathcal{E}) < 0$. Consider the following choice set $\mathcal{S}$ (with payoffs in utils):
\begin{align*}
 & \texttt{safe}: \{ 0 \textrm{ always} \}, \\
 & \texttt{risky}: \{ a \textrm{ if $A \cap E$}, -b \textrm{ if $A^{C} \cap E$}, 0 \textrm{ otherwise} \}.
\end{align*}
We want to find values for $a > 0$ and $b > 0$ such that, conditional on $E$, the expected utility of the risky bet is worse:
\begin{equation}\label{constraint1}
\mathbb{E}_{p(\cdot \mid E)}(\texttt{risky}) < 0.
\end{equation}
But if our agent deviates from conditionalization, she prefers the risky bet:
\begin{equation}\label{constraint2}
\mathbb{E}_{p_{E}}(\texttt{risky}) > 0.
\end{equation}
If we find these values, we can show that $Val_{General}(\mathcal{E}) < 0$. Recall that 
\begin{equation*}
Val_{General}(\mathcal{E}) =  \mathbb{E}_{p} \left( \argmax_{f \in \mathcal{S}} \mathbb{E}_{\mathcal{P}_{\mathcal{E}}} (f) \right)
 - \max_{f \in \mathcal{S}} \mathbb{E}_{p} (f).
\end{equation*}
Now $\max_{f \in \mathcal{S}} \mathbb{E}_{p} (f) = 0$. This is because $\mathbb{E}_{p} ( \texttt{safe}) = 0$ but $\mathbb{E}_{p} ( \texttt{risky}) < 0$. We need to show 
\begin{equation}\label{negativeex}
\mathbb{E}_{p} \left( \argmax_{f \in \mathcal{S}} \mathbb{E}_{\mathcal{P}_{\mathcal{E}}} (f) \right) <  0.
\end{equation}
We re-write this term using the law of total expectation: 
\begin{equation*}
p(E)\mathbb{E}_{p(\cdot \mid E)} \left( \argmax_{f \in \mathcal{S}} \mathbb{E}_{\mathcal{P}_{E}} (f) \right) + p(E^C) \mathbb{E}_{p(\cdot \mid E^C)} \left( \argmax_{f \in \mathcal{S}} \mathbb{E}_{\mathcal{P}_{E^C}} (f) \right).
\end{equation*}
Now the right-hand term is zero, since both \texttt{safe} and \texttt{risky} yield zero when $E$ is false. Thus, we focus on the left-hand term, which we can re-write as follows, using Evidential Independence and Lemma (\ref{lemma1}):
\begin{equation*}
p(\textrm{choose \texttt{risky}} \mid E) \mathbb{E}_{p(\cdot \mid E)} ( \texttt{risky}) + p(\textrm{choose \texttt{safe}} \mid E) \mathbb{E}_{p(\cdot \mid E)} ( \texttt{safe}).
\end{equation*}
The right-hand term is again zero, so we focus on the left-hand term. We have $p(\textrm{choose \texttt{risky}} \mid E) > 0$, since we have assumed that there is a positive probability our agent deviates from conditionalization and so chooses the risky action. By assumption, $\mathbb{E}_{p(\cdot \mid E)} ( \texttt{risky}) < 0$, which shows (\ref{negativeex}).

We still need to show that we can find values $a$ and $b$ which do the trick. By (\ref{constraint1}), $a$ and $b$ need to obey the following constraint:
\begin{equation*}
ap(A \cap E \mid E) - bp(A^{C} \cap E\mid E) < 0,
\end{equation*}
which simplifies to
\begin{equation}
ap(A \mid E) - b (1-p(A \mid E)) < 0.
\end{equation}
By (\ref{constraint2}), $a$ and $b$ need to obey the following constraint:
\begin{equation*}
a p_{E}(A \cap E) - b p_{E}(A^{C} \cap E) > 0,
\end{equation*}
which, using our assumption that $p_{E}(E)=1$, simplifies to
\begin{equation}
a p_{E}(A) - b (1 - p_{E}(A)) > 0.
\end{equation}
Let us write $q$ for $p_{E}(A)$ and $r$ for $p(A \mid E)$. So our question is whether the following system of equations has a solution for any $q$ and $r$ such that $q > r$:
\begin{equation}
aq - b(1-q) > 0 > ar - b(1-r).
\end{equation}
The answer is `yes': real numbers $a > 0$ and $b >0$ such that $0 \leq r < \frac{b}{a+b} < q \leq 1$. We can find outcomes with these utilities by Utility Richness.
\end{proof}

\begin{theorema}
Assuming Evidential Independence, $\mathcal{E}$, $Val_{General}(\mathcal{E}) \leq Val_{Good}(\mathcal{E})$ for every evidence partition $\mathcal{E}$.
\end{theorema}

\begin{proof}
It suffices to show
\begin{equation}\label{inequal2}
\sum_{E \in \mathcal{E}} p(E) \mathbb{E}_{p(\cdot \mid E)} \left( \argmax_{f \in \mathcal{S}} \mathbb{E}_{\mathcal{P}_{E}} (f)  \right) \leq \mathbb{E}_p \left( \max_{f \in \mathcal{S}} \mathbb{E}_{\mathcal{P}(\cdot \mid \mathcal{E})} (f) \right).
\end{equation}
Consider any $E \in \mathcal{E}$. By Evidential Independence and Lemma (\ref{lemma1}),
\begin{equation}\label{inequal3}
 \mathbb{E}_{p(\cdot \mid E)} \left( \argmax_{f \in \mathcal{S}} \mathbb{E}_{\mathcal{P}_{E}}(f) \right) = \sum_{i=1}^n p(\textrm{choose $f_i$} \mid E) \mathbb{E}_{p(\cdot \mid E)} (f_i).
\end{equation}
The right-hand side is a weighted sum of expected values relative to $p(\cdot \mid E)$ and $\mathbb{E}_{p(\cdot \mid E)} (f) \leq \max_{f \in \mathcal{S}} \mathbb{E}_{p(\cdot \mid E)} (f)$ for all $f \in S$.\footnote{This is a version of the principle that, for expected utility maximizers, randomization can never be strictly preferable \parencite[119-120]{Icard2021}. In general, this follows from Jensen's inequality.} Therefore,
\begin{equation*}\label{inequal4}
 \sum_{i=1}^n p(\textrm{choose $f_i$} \mid E) \mathbb{E}_{p(\cdot \mid E)} (f_i) \leq \max_{f \in \mathcal{S}} \mathbb{E}_{p(\cdot \mid E)} (f),
\end{equation*}
and so by (\ref{inequal3}),
\begin{equation*}\label{inequal5}
 \mathbb{E}_{p(\cdot \mid E)} \left( \argmax_{f \in \mathcal{S}} \mathbb{E}_{\mathcal{P}_{E}}(f) \right) \leq \max_{f \in \mathcal{S}} \mathbb{E}_{p(\cdot \mid E)} (f).
\end{equation*}
Taking expectations on both sides, (\ref{inequal2}) follows.
\end{proof}

\section*{Acknowledgements}

Thanks to Lara Buchak, Kevin Dorst, James Evershed, Wesley Holliday, Mikayla Kelley, John MacFarlane, Milan Moss\'{e}, Richard Pettigrew, Rebecca Rowson, David Thorstad, Yong Xin Hui, Snow Zhang, and two anonymous referees for very helpful comments on earlier drafts. Also thanks to Mathias B\"{o}hm, Morgan Connolly, Johann Frick and Edward Schwartz for insightful discussions and to audiences at Berkeley's awesome Formal Epistemology Reading Course (FERC), `Bayesian Epistemology: Perspectives and Challenges' at the Munich Center for Mathematical Philosophy, Inquiry Network, Formal Ethics 2022, Berkeley's Richard Wollheim Society, Central APA 2022, the Berkeley-London Conference 2022 and the Bochum Bayesian Learning and Reasoning Workshop 2022. During research, I was supported by a Global Priorities Fellowship by the Forethought Foundation and a Josephine De Karman Fellowship.

\printbibliography

\end{document}